\documentclass[]{aastex63}
\usepackage{mhchem}

\submitjournal{ApJ}
\shortauthors{Hongxuan et al.}

\begin{document}

\title{Photosphere Recession and Luminosity of Homologous Explosions Revisited}
\correspondingauthor{Xue-Wen Liu}
\email{liuxuew@scu.edu.cn}

\author{Hong-Xuan Jiang}
\affiliation{College of Physics, Sichuan University, 610065, Chengdu, China}
\affiliation{Tsung-Dao Lee Institute, Shanghai Jiao-Tong University, 200240, Shanghai, China}
\author{Xue-Wen Liu}
\affiliation{College of Physics, Sichuan University, 610065, Chengdu, China}

\author{Zhi-Yong You}
\affiliation{Key Laboratory of Particle Astrophyics \& Experimental Physics Division \& Computing Center,
Institute of High Energy Physics, Chinese Academy of Sciences, 100049 Beijing, China}
\affiliation{University of Chinese Academy of Sciences, 100049 Beijing, China}
\affiliation{TIANFU Cosmic Ray Research Center, Chengdu, Sichuan, China}




\begin{abstract}
  By assuming the photosphere located at the outmost edge of the ejecta,
  Arnett et al. (1980, 1982, 1989) presented the light curves of homologous
  explosions in supernovae analytically and numerically to include recombination effects.
  Actually as homologous expansion proceeds, the photosphere recedes deeper
  into the ejecta. In this situation, the photosphere radius increases at early times
  and decreases later on which can be described by a simple method proposed by Liu et al. (2018).
  To study how the photosphere recession effect the luminosity evolution,
  we impose a boundary condition on the photosphere to determine the spatial and time distribution of
  the temperature of the ejecta which is clarified to be reasonable.
 We find that the photosphere recession reduce the luminosity compared with the previous result without the
  recession, which can be tested with observations of Type-IIP supernovae.
\end{abstract}



\section{Introduction} \label{sec:intro}
The analytical and semianalytical light curve models allow crude and quick estimates of explosion energy, initial radius, the mass of ejected matter
and $^{56}{\text{Ni}}$ to be made from the observed data of supernovae. All these models assume a stationary
photosphere fixed at the outmost edge of the ejecta
(\citep{1980ApJ...237..541A, 1982ApJ...253..785A, 1989ApJ...340..396A}; \cite{Nagy2014}). Because of the continuous expansion of the ejecta,
the photosphere radius always increases. But the photosphere radius
$R_{\text{ph}}(t)=[L_{\text{bol}}(t)/4\pi \sigma T^4_{\text{eff}}(t)]^{1/2}$ of
some supernovae with well observed bolometric luminosity $L_{\text{bol}}$ and effective
temperature $T_{\text{eff}}$ show an early rising and late falling behavior. To explain
this evolution, Liu et al. (2018) allows the photosphere recede within the ejecta
as its expansion and derive an ordinary differential equation to describe this recession.
Although indeed showing a general rising/falling behavior, their results cannot fit the
data quite well. One possibility is the complex relation between the photosphere radius and
the bolometric luminosity not as $R_{\text{ph}}\propto L_{\text{bol}}^{1/2}$. For the photosphere fixed at the edge of the envelope, its radius is easy to get. \cite{1980ApJ...237..541A, 1982ApJ...253..785A} use the Eddington surface boundary condition on the fixed photosphere to determine the spatial and time structure of the temperature below the photosphere and then obtain the luminosity. In this case, an eigenvalue $\alpha$ in separating the spatial and time structures governed by partial differential equations is found to be constant. However time-dependent $\alpha$ is necessary when the recombination effect is included \citep{1989ApJ...340..396A,1993ApJ...414..712P}. The recombination front moving through the envelope plays a role as the pseudo photosphere. Based on
\cite{1989ApJ...340..396A}, \cite{Nagy2014} added the magnetar as an extra energy input to explain the Type-IIP supernovae. These previous works didn't consider the photosphere recession effect. Actually, the photosphere will recede into the envelope as its expansion in any case.
In this paper, we impose the Eddington boundary condition on the receding photosphere to determine its position and then
study the luminosity evolution. We assume a homologous expansion and spherically symmetric
supernovae ejecta as the same as previous papers did. And we also treat the radiation transport by
the diffusion approximation.

This paper is organized as follows: in Section 2 we obtain the radius evolution of the receding
photosphere in a different way. In Section 3 we derive the luminosity evolution. The results are
summarised in Section 4. Finally, we give a summary in Section 5.

\section{photosphere radius evolution} \label{sec:photosphere}


The radius of the photosphere in the ejecta is determined by
\begin{equation}
  \int_{R_{\text{ph}}\left(t\right)}^{R\left(t\right)} \rho\left(r, t\right) \kappa dr  = \frac{2}{3}, \label{Eq: R_{ph}}
\end{equation}
where $\rho\left(r, t\right)$ is the density of the ejecta, $\kappa$ is the opacity and $R\left(t\right)$ is the surface radius of the ejecta (Arnett 1980).
Because Thomson scattering dominates the opacity, we use a constant $\kappa$ throughout the evolution. In the coasting phase, the surface radius of the homologous expanding ejecta is
\begin{equation}
  R(t)=R(0)+v_{\text{sc}}t,
\end{equation}
where $R(0)$ is the initial surface radius and $v_{\text{sc}}$ is the velocity scale.
The density can be separated into
\begin{equation}
  \rho\left(r,t\right)=\frac{\rho\left(0,0\right)\eta\left(x\right)R\left(0\right)^3}{R\left(t\right)^3}, \label{Eq: rho}
\end{equation}
where $0 \leq x \leq 1$ is a dimensionless radius $x=r/R(t)$, $\rho(0,0)$ is the initial density at the innermost edge and is written as $\rho_0$ hereafter. 



From Eq.(\ref{Eq: R_{ph}}) and Eq.(\ref{Eq: rho}), we obtain an integration including the dimensionless
photosphere radius
\begin{equation}
\int_{x_{\text{ph}}\left(t\right)}^1 \rho_0 \eta\left(x\right)\frac{R(0)^3}{R\left(t\right)^2}\kappa dx	= \frac{2}{3}. \label{Eq: x_{ph}}
\end{equation}
As long as $\eta\left(x\right)$ is given, the evolution of the receding photosphere can be obtained.
We use a Paczy\'nski red supergiant envelope as an example, i.e., $\eta(x)=e^{A x}$ with $A = -1.732$.
Substitute it into Eq.(\ref{Eq: x_{ph}}) and we obtain
\begin{equation}
  x_{\text{ph}}(t) = \frac{\ln\left[e^A-\frac{2A\left(R(0) + v_{sc} t\right)^2}{3R(0)^3\kappa\rho_0} \right]}{A}, \label{Eq: expression of xph}
\end{equation}
which is the same as the result of \cite{2018ApJ...868L..24L}.

%

\section{Luminosity Evolution}\label{Sec: light curve}
In this section we briefly describe the Arnett's model \citep{1980ApJ...237..541A, 1982ApJ...253..785A}
firstly. It assumed that photons come from the surface of the ejecta,
which is the photosphere in their model. In fact as its expanding, the ejecta becomes thinner
and the photosphere will naturally recedes into the inner part of the ejecta.
So next we propose a method to determine the position of the receding photosphere and get the solution
of the luminosity.


\subsection{Arnett's model of supernova} \label{Subsec: Arnett}
In the diffusion approximation, the luminosity is 
\begin{equation}
  L/4\pi r^2=-(\lambda c/3)\partial aT^4/\partial r, \label{Eq: diffusion approximation}
\end{equation}
where the mean free path is $\lambda = 1/(\rho \kappa)$ and the mass density is $\rho=1/V$.
In the strictly adiabatic case, the temperature can be separated as 
\begin{equation}
  T\left(r, t\right)^4=\frac{\psi\left(x\right)\phi\left(t\right)T\left(0,0\right)^4 R\left(0\right)^4}{R\left(t\right)^4}. \label{Eq: T^4}
\end{equation}  
Thus the photosphere luminosity can be written as 
%
%
%
\begin{equation}
  L\left(x_{\text {ph}}, t\right) =-\frac{4\pi a T\left(0, 0\right)^4c R(0)}{3\rho\left(0,0\right)\kappa}\phi\left(t\right)\left(\frac{x^2}{\eta(x)}\frac{d\psi\left(x\right)}{dx}\right)\bigg{|}_{x=x_\text{ph}}, \label{Eq: Luminosity}
\end{equation}
where $\phi(t)$ can be solved by the thermodynamics of the trapped radiation in the expansion ejecta and $\left(\frac{x^2}{\eta(x)}\frac{d\psi\left(x\right)}{dx}\right)\bigg{|}_{x=x_\text{ph}}$ is determined by the
boundary condition. According to the first law of thermodynamics, the thermal state of the expanding matter
evolves in time as 
\begin{equation}
  \dot E+P\dot V=-\frac{\partial L}{\partial m} + \varepsilon, \label{Eq: first law of thermodynamics}
\end{equation} 
where $E$ is the thermal energy per unit mass, $P$ is the pressure and $\varepsilon$ is
the energy release per unit mass from radioactive decay.
For a radiation dominated gas, the energy and pressure are $E=aT^4V$ and $P=aT^4/3$. 
Substitute Eq.(\ref{Eq: T^4}) and Eq.(\ref{Eq: diffusion approximation})
into Eq.(\ref{Eq: first law of thermodynamics}) and note that
$\partial L/\partial m=(1/4\pi r^2 \rho)\partial L/\partial r$, Eq.(\ref{Eq: first law of thermodynamics})
becomes
\begin{equation}
  aT^4 V \frac{\dot \phi(t)}{\phi(t)}=\varepsilon+\frac{1}{r^2 \rho}\frac{\partial}{\partial r}\left(\frac{cr^2}{3\kappa\rho} \frac{\partial aT^4}{\partial r}\right). \label{Eq: phidotphi}
\end{equation}
Now as in A80, let
\begin{equation}
  V(r,t)=V(0,0)[R(t)/R(0)]^3/\eta(x),
\end{equation}
and
\begin{equation}
  \varepsilon=\varepsilon_{\text{Ni}}^0\xi(x)e^{-t/\tau_{\text{Ni}}}.
\end{equation}
%
Using these expressions, Eq.(\ref{Eq: phidotphi}) reduces to
\begin{equation}
  \frac{R(0)}{R(t)}\frac{\dot \phi(t)}{\phi(t)}- \bigg[\frac{\varepsilon_{\text{Ni}}^0}{ aT\left(0,0\right)^4V(0,0)}\bigg]\bigg[\frac{ \xi(x)\eta(x)}{\psi(x)}\bigg]\frac{e^{-t/\tau_{\text{Ni}}}}{\phi(t)}=-\frac{cV(0,0)}{3\kappa R(0)^2}\frac{1}{\psi(x) x^2} \frac{\partial}{\partial x}\left[\frac{x^2}{\eta(x)}\frac{\partial \psi}{\partial x}\right]. \label{Eq: alpha_time}
\end{equation}
%
As in A82, if we assume that 
\begin{equation}
  b\equiv\frac{\eta(x)\xi(x)}{\psi(x)}
\end{equation}
is constant for any $x$.
%
%
It's obvious that Eq.(\ref{Eq: alpha_time}) is separable.
\cite{1982ApJ...253..785A} defined two parameters
\begin{equation}
  \alpha\equiv -\frac{1}{\psi(x)x^2}\frac{\partial}{\partial x}\left[\frac{x^2}{\eta(x)}\frac{\partial \psi}{\partial x}\right], \label{Eq: alpha_space}
\end{equation} 
and
\begin{equation}
  \tau_0=\frac{3\kappa R(0)^2}{\alpha V(0,0) c}. \label{Eq: tau0}
\end{equation}
Finally, the time evolution of $\phi(t)$ is governed by Eq.(\ref{Eq: alpha_time}) which becomes
\begin{equation}
  \dot \phi(t) + \frac{R(t)}{R(0)\tau_0}\phi(t) = \bigg[\frac{b\varepsilon_{\text{Ni}}^0}{ aT\left(0,0\right)^4V(0,0)}\bigg]\bigg[\frac{R(t)}{R(0)}\bigg]e^{-t/\tau_{\text{Ni}}}. \label{Eq: phit}
\end{equation}
%
This time solution and the spatial solution of Eq.({\ref{Eq: alpha_space}}) can be solved if the density profile $\eta(x)$ and the spacial boundary condition are given. Then the luminosity of Eq.(\ref{Eq: Luminosity}) will be obtained.

\subsection{Boundary Conditions}
As in A80, at the center we can impose the initial conditions $d\psi(x=0)/dx=0$ and $\psi(0)\equiv 1$ to get
the solution of Eq.($\ref{Eq: alpha_space}$) which is a function of the parameter $\alpha$. The previous literature
use a "radiative-zero" boundary condition $\psi(1)=0$ to get the value of $\alpha$. But this boundary condition
is only valid for dense objects. In this paper, we consider that the photosphere recedes into the envelope as its expanding and getting thinner.
So the boundary changes as expansion, which will produce a time dependent $\alpha$.

According to \cite{1926ics..book.....E}, we have
\begin{equation}
  acT^4=H(2+3\tau),
\end{equation}
where $H=\frac{1}{4\pi}\int J(\theta) \cos{\theta} d\omega$. When $\tau$ equals to $2/3$ the envelope becomes transparent for photons. So at the position of photosphere we can obtain
\begin{equation}
  acT_{\text{ph}}^4=4H.
\end{equation}
And thus
\begin{equation}
  T^4=\frac{1}{4}T_{\text{ph}}^4(2+3\tau).
\end{equation}
Since $T^4 \propto \psi$, we can obtain
\begin{equation}
  \psi=\frac{1}{4}\psi_{\text{ph}}(2+3\tau).
\end{equation}
Take the derivative of both sides, we get
\begin{equation}
  \frac{\partial \psi(x)}{\partial x}=\frac{3}{4}\psi_{\text{ph}}\frac{\partial \tau}{\partial x}=-\frac{3}{4}\psi_{\text{ph}}\frac{R(t)}{\lambda(x)}.
\end{equation}
Then the boundary condition of the photosphere gives
\begin{equation}
  \psi(x_{\text{ph}})=-\frac{4}{3}\frac{\lambda(x_{\text{ph}})}{R(t)} \frac{\partial \psi(x)}{\partial x}\bigg|_{x = x_{\text{ph}}}. \label{Eq: boundary condition}
\end{equation}
For a dense object, $\lambda(x_{\text{ph}})/R(t) \ll 0$, so $\psi(x_{\text{ph}}\to 1)\to 0$ which is just the "radiative-zero" boundary condition $\psi(1)=0$. 
However, in our treat the outside of the photosphere is no longer dense but thin enough to be optical transparent. Then the ``radiative-zero boundary condition is no longer valid. The boundary condition of Eq.(\ref{Eq: boundary condition}) mean that the photosphere is no longer fixed but time dependent which leads to a time dependent $\alpha$.
We stress that all the differences come from the photosphere recession. The term $R(0)^3/R(t)^3$ in Eq.(\ref{Eq: rho}) term makes the density thinner and thinner so that the photosphere will naturally recede. In this case, $\psi(x)$ is also time dependent. At any moment, we always have Eq.(\ref{Eq: alpha_space}) and Eq.(\ref{Eq: alpha_time}) established. We just need to make $\psi(x)$ satisfy the boundary condition of Eq.(\ref{Eq: boundary condition}) during the recession. And this method is also used in \cite{1993ApJ...414..712P} and \cite{1989ApJ...340..396A}. They use a time-dependant $\alpha$ to identify the position of the moving recombination front. 

\subsubsection{Luminosity with Fixed Boundary}
To simplify the luminosity of Eq.(\ref{Eq: Luminosity}) further. A82 defined
the initial mass of \ce{^{56}Ni} in the envelope
\begin{equation}
  M_{\text{Ni}}^0=\frac{4\pi R(0)^3}{V(0,0)}\int_0^1\eta(x)\xi(x)x^2dx, \label{Eq: MNi}
\end{equation}
and the initial internal energy
\begin{equation}
  E_{\text{Th}}^0 = 4\pi R(0)^3 aT(0,0)^4\int_0^1\psi(x)x^2 dx, \label{Eq: Eth}
\end{equation}
where
\begin{equation}
  I^0_{\text{Th}} \equiv \int_0^1\psi(x)x^2dx. \label{Eq: Ith}
\end{equation}
Since
\begin{equation}
  b\equiv\frac{\eta(x)\xi(x)}{\psi(x)}=\frac{\int_0^1\eta(x)\xi(x)x^2dx}{\int_0^1\psi(x)x^2dx},\label{Eq: bb}
\end{equation}
we use Eq.(\ref{Eq: MNi}), Eq.(\ref{Eq: Eth}) and Eq.(\ref{Eq: bb}) to reduce Eq.(\ref{Eq: phit}) to
\begin{equation}
  \dot \phi = \left[\frac{\varepsilon_{Ni}^0 M_{\text{Ni}}^0}{E_{\text{Th}}^0}e^{-t/\tau_{\text{Ni}}}-\frac{\phi}{\tau_0}\right]\frac{R(t)}{R(0)}. \label{Eq: phi constant alpha}
\end{equation}
Integrating Eq.(\ref{Eq: alpha_space}) by part and evaluating at $x=0,1$ gives
\begin{equation}
  \alpha\int_0^1 \psi(x)x^2dx=-\left(\frac{x^2}{\eta(x)}\frac{d\psi\left(x\right)}{dx}\right)\bigg{|}_{x=1}. \label{Eq: fixedboundary}
\end{equation}
Using Eq.(\ref{Eq: Eth}), Eq.(\ref{Eq: fixedboundary}) and Eq.(\ref{Eq: tau0}), the luminosity of Eq.(\ref{Eq: Luminosity}) reduces to a very simple expression
\begin{equation}
  L(1,t)=\frac{E_{\text{Th}}^0}{\tau_0}\phi(t),\label{Eq: LumFix}
\end{equation}
which can be solved analytically (see A82).

\subsubsection{Luminosity from Receding Photosphere}
The observed emission comes from the photosphere. The first law of thermodynamics of Eq.(\ref{Eq: first law of thermodynamics}) within the volume enclosed by the photosphere gives
\begin{equation}
  \dot E_{\text{ph}}+(P\dot V)_{\text{ph}}=-\frac{\partial L_{\text{ph}}}{\partial m}+\varepsilon_{\text{ph}},
  \label{Eq: firstlawEph}
\end{equation}
where the internal energy enclosed by the photosphere is
\begin{equation}
  E_{\text{ph}}=\int_0^{R_{\text{ph}}}4\pi r^2 aT^4dr=4\pi R(0)^3 aT(0,0)^4
  \phi(t)\frac{R(0)}{R(t)}\int_0^{x_{\text{ph}}}\psi(x)x^2dx\label{Eq: Eph}
\end{equation}
By defining a new parameter
\begin{equation}
  I_{\text{ph}}=\int_0^{x_{\text{ph}}}\psi(x)x^2dx,\label{Eq: Iph}
\end{equation}
Eq.(\ref{Eq: Eph}) is rewritten as
\begin{equation}
  E_{\text{ph}}=E_{\text{Th}}^0\frac{I_{\text{ph}}}{I^0_{\text{Th}}}\frac{R(0)}{R(t)}\phi(t).
\end{equation}
Therefore
\begin{equation}
\begin{aligned}
  \dot E_{\text{ph}} = &E_{\text{ph}}\left(\frac{\dot \phi(t)}{\phi(t)}-\frac{v_{\text{sc}}}{R(t)} +
    \frac{\dot I_{\text{ph}}}{I_{\text{ph}}} \right)\\
  = &E_{\text{ph}}\left(\frac{d\ln{\phi(t)}}{dt}-\frac{d\ln R(t)}{dt} + \frac{d\ln{I_{\text{ph}}}}{dt} \right). \label{Eq: left side}
\end{aligned}
\end{equation}
Using $d\ln V/dt=3d\ln R(t)/dt$, the pressure $P=E/3$ and Eq.(\ref{Eq: left side}) to
eliminate $d\ln R(t)/dt$ from Eq.(\ref{Eq: firstlawEph}) gives
\begin{equation}
  E_{\text{ph}}\left(\frac{\dot \phi(t)}{\phi(t)} +
    \frac{\dot I_{\text{ph}}}{I_{\text{ph}}} \right)= -\frac{\partial L}{\partial m}\bigg |_{x_{\text{ph}}}+\varepsilon.
\end{equation}
Here we assume a central energy production and therefore $\varepsilon_{\text{ph}}=\varepsilon$. According to the
definition of Eq.(\ref{Eq: Iph}), the time derivative of $I_{\text{ph}}$ is
\begin{equation}
  \dot I_{\text{ph}} = \psi(x_{\text{ph}})x_{\text{ph}}^2\dot x_{\text{ph}}.
\end{equation}
Now as the same procedures as presented in Sec.(\ref{Subsec: Arnett}), the temporal part can be obtained as follows:
\begin{equation}
  \frac{d \phi(t)}{dt} =\frac{R(t)}{R(0)}\left[p_1 \frac{I_{\text{th}}}{I_{\text{ph}}}e^{-t/\tau_{\text{Ni}}}-\frac{\phi(t)}{\tau_0}\right]-
  \phi(t)\frac{\psi(x_{\text{\text{ph}}})x_{{\text{ph}}}^2}{I_{\text{ph}}}\frac{dx_{\text{ph}}}{dt} ,\label{Eq: phi}
\end{equation}
where $p_1 = \varepsilon^0_{\text{Ni}}M_{\text{Ni}}^0/E_{\text{Th}}^0$. Integrating Eq.(\ref{Eq: alpha_space}) by part and evaluating at $x=0,x_{\text{ph}}$ gives
\begin{equation}
  \left(\frac{x^2}{\eta(x)}\frac{d\psi\left(x\right)}{dx}\right)\bigg{|}_{x=x_{\text{ph}}}=-\alpha\int_0^{x_{\text{ph}}} \psi(x)x^2dx. 
\end{equation}
The luminosity from receding photosphere is therefore written as
\begin{equation}
  L(x_{\text{ph}},t)=\frac{E^0_{\text{Th}}}{\tau_0}\frac{I_{\text{ph}}}{I^0_{\text{Th}}}\phi(t), \label{Eq: LumRec}
\end{equation}
which can only be solved numerically.

\subsubsection{Luminosity from Receding Photosphere: broken power law density profile}
To understand how different environment effect the recession, we also consider the broken power law density profile which is usually assumed as follows \citep{1982ApJ...258..790C, 1999ApJ...510..379M,2010ApJ...717..245K, 10.1093/mnras/stt1392}
\begin{equation}
	\eta(x)=\left\{
\begin{aligned}
& \left(x/x_0\right)^{-\delta} \, 0\leq x\leq x_0, \\
& \left(x/x_0\right)^{-n} \, x_0\leq x\leq 1,
\end{aligned}
\right.
\end{equation}
where $x_0$ is a dimensionless transition radius that divide the inner part and the outer part of the envelope. For SN Ib/Ic and SN Ia progenitors, one has $n\sim10$ \citep{1999ApJ...510..379M, 10.1093/mnras/stt1392}. We choose the parameters of $x_0=0.1$, $\delta = 0$ and $n=10$ as the same as in \cite{2016A&A...589A..53N}. We can express $x_{\text{ph}}$ analytically
\begin{equation}
  x_{\text{ph}}(t)=\left\{
    \begin{aligned}
      & \left(\frac{(1-n) x_0^{-n} R^2(t) \left(-\frac{\kappa  \rho_0 R\left(0\right)^3 x_0^n }{(n-1) R^2(t)}-\frac{2}{3}\right)}{\kappa  \rho_0 R\left(0\right)^3}\right)^{\frac{1}{1-n}}, \, x_{\text{ph}}\textgreater x_0, \\
      & x_0 -\frac{2R^2\left(t\right)}{3 R\left(0\right)^3 \rho_0 \kappa}, \, x_{\text{ph}}\leq x_0,
    \end{aligned}
  \right.
\end{equation}
In this case, the density in the outer part of the envelope is significantly lower which causes the temperature profile $\psi(x)$ in that region is very close to zero. Therefore the gradient of density and temperature is almost zero. According to Eq. (\ref{Eq: boundary condition}), the parameter $\alpha$ is highly related to the temperature gradient. Thus $\alpha$ doesn't change much and as a result the luminosity in our model is quite similar to the former work as shown in Fig. 1.

\subsubsection{Luminosity from Receding Photosphere: constant density profile}
The constant density profile is widely used in the research of Type-IIP supernovae \citep{Nagy2014, 2003MNRAS.338..711Z, Chatzopoulos2012}. We let $\eta(x) = 1$ to obtain
\begin{equation}
  x_{\text{ph}}=1-\frac{2R^2\left(t\right)}{3 R\left(0\right)^3 \rho_0 \kappa}.
\end{equation}
As shown in Fig. 1, the luminosity decreases much faster with higher density in the outer region of the envelope. Such result is understandable. In the broken-power-law case, the whole envelope is like a dense core with a thin shell, the shell is almost transparent. Basically we are observing the inner part of the envelope into which the photosphere doesn't recede very deeper. In the exponential and constant case the temperature and density gradient is higher, thus the photosphere recession effect is stronger and causes the luminosity obvious lower than the fixed photosphere model.

\section{The Results and Discussions}
We revise the luminosity evolution of homologous explosion by considering the photosphere
  recession. Now we compare our numerical results with the fixed photosphere model.
We choose the parameters of $R(0)=1.59\times 10^{14}\text{cm}$, $M=10M_\sun$,
$\kappa=0.33\ \text{cm}^2 g^{-1}$, $v_{\text{sc}}=1.0\times 10^9\ \text{cm}\ s$,
$E_{\text{Th}}^0=2.0\times 10^{51}\ \text{ergs}$ and adopt three different envelope environments. To see
clearly the recession effect, we don't include the energy source of radioactive decay from $^{56}{\text{Ni}}$ which
just produces an exponential tail at late time.
In Fig. 1, the luminosity with fixed and recession photosphere is presented. It is shown that the
photosphere recession reduce the luminosity compared with the results from the photosphere
fixed at the surface. We find that the density gradient is the most important factor to reduce
the luminosity. 

Our luminosity formula of Eq.(\ref{Eq: LumRec}) is general. The previous result of Eq.(\ref{Eq: LumFix})
is just a special case. The ratio $I_{\text{ph}}/I_{\text{Th}}^0$ between Eq.(\ref{Eq: LumRec}) and
Eq.(\ref{Eq: LumFix}) is equal to 1 for $x_{\text{ph}}=1$, which means that our formula recover the previous fixed photosphere result if the photosphere does not recede. We show the time evolution of the ratio for the exponential density environment in Fig .2. The ratio reduction due to the recession is obvious. In Fig.2, we also
  show the numerical solution of $\phi(t)$ which exhibits a late decrease too. The two factors combine to result in the whole behaviour of the luminosity. Although photosphere recession is newly considered, we would like to stress that the behind idea introducing the time dependent $\alpha$ was already used in \cite{1989ApJ...340..396A} and \cite{1993ApJ...414..712P} to locate the position of the recombination front. While they used this method to include the recombination effect to explain the plateau of Type-IIP supernovae, the photosphere is still fixed at the surface, not at the recombination front. In the same way, we determine the position of the receding photosphere from which the luminosity radiate. The receding photosphere is a real photosphere not like the recombination front. 
 We can furthermore conjecture that the photosphere should recede even the recombination of hydrogen takes place, which beyond the scope of this paper. 
  
The decrease of luminosity caused by the photosphere recession is significant at late times as shown in Fig 1. In this period, the luminosity due to the nebula or the magnetar  becomes to exceed the photosphere emission \citep{Wang_2016a}. Therefor it is difficult to observe the recession directly. However, we find that the behavior of the luminosity considering the recession in about the initial 100 days after the burst is quite different from previous results if the recombination effect is included. The recombination of hydrogen highly effect 
the light curves of Type-II supernovae, especially for Type-IIP supernovae.

In summary, we solve the luminosity of homologous explosion in supernovae
considering the photosphere recession for the first time. Fitting supernovae
and other optical transients data in our future work (in preparation) will be used to
find the evidence of the photosphere recession.


%
\begin{figure}[ht!]
  \plotone{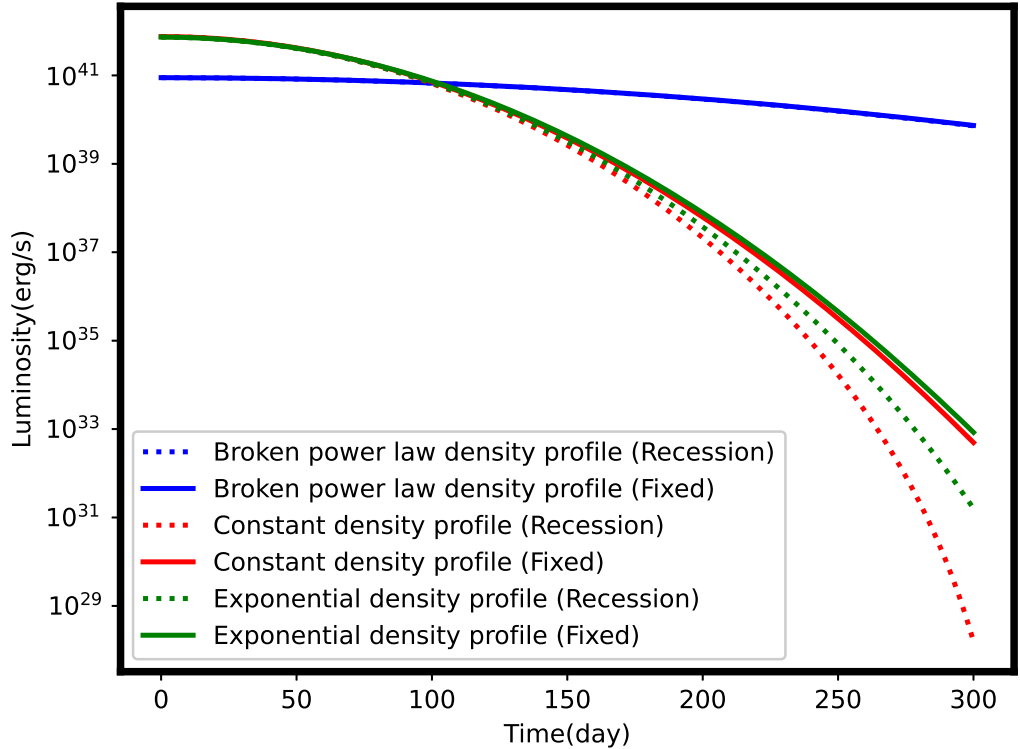}
\caption{Luminosity of fixed and recession photosphere from three different density environments.}
\end{figure} 
\begin{figure}[ht!]
  \plotone{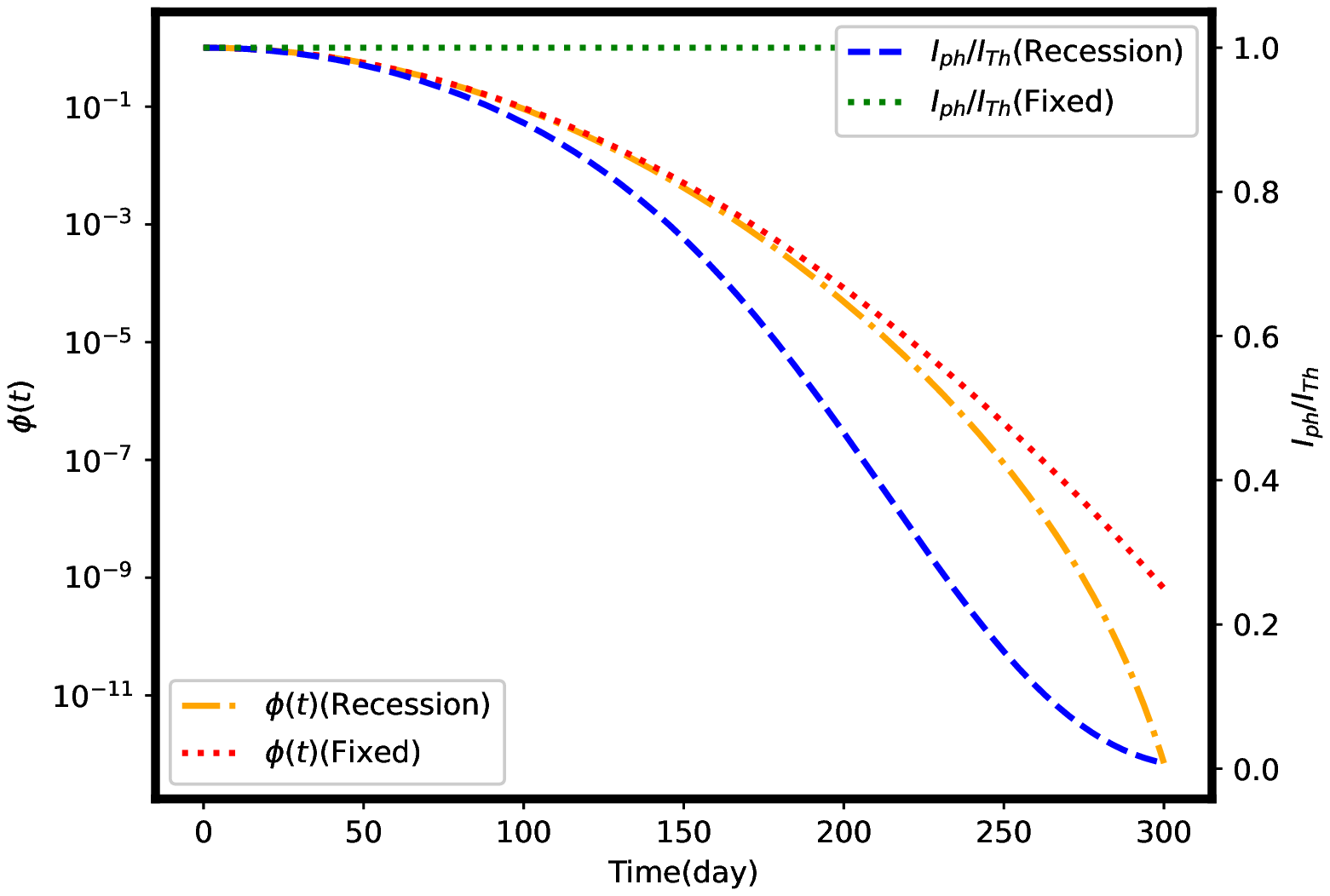}
\caption{Evolution of $\phi(t)$ and $I_{\text{ph}}/I^0_{\text{Th}}$ for fixed and recession photosphere for the exponential density environment.}
\end{figure} 
\begin{figure}[ht!]
\plotone{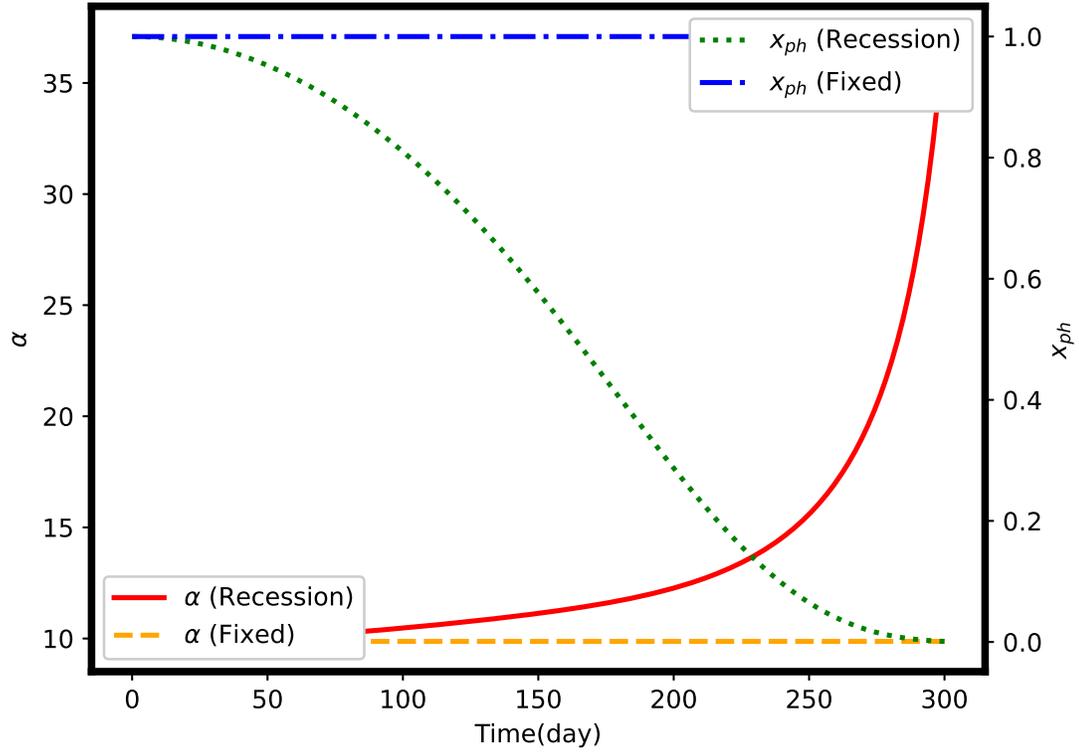}
\caption{Evolution of $\alpha$ and $x_{\text{ph}}$ for fixed and recession photosphere for the exponential density environment.}
\end{figure} 

\bibliographystyle{aasjournal}

\begin{thebibliography}{}
\expandafter\ifx\csname natexlab\endcsname\relax\def\natexlab#1{#1}\fi
\providecommand{\url}[1]{\href{#1}{#1}}
\providecommand{\dodoi}[1]{doi:~\href{http://doi.org/#1}{\nolinkurl{#1}}}
\providecommand{\doeprint}[1]{\href{http://ascl.net/#1}{\nolinkurl{http://ascl.net/#1}}}
\providecommand{\doarXiv}[1]{\href{https://arxiv.org/abs/#1}{\nolinkurl{https://arxiv.org/abs/#1}}}

\bibitem[{{Arnett}(1980)}]{1980ApJ...237..541A}
{Arnett}, W.~D. 1980, \apj, 237, 541, \dodoi{10.1086/157898, A80}

\bibitem[{{Arnett}(1982)}]{1982ApJ...253..785A}
---. 1982, \apj, 253, 785, \dodoi{10.1086/159681,A82}

\bibitem[{{Arnett} \& {Fu}(1989)}]{1989ApJ...340..396A}
{Arnett}, W.~D., \& {Fu}, A. 1989, \apj, 340, 396, \dodoi{10.1086/167402, A89}

\bibitem[{Chatzopoulos {et~al.}(2012)Chatzopoulos, Wheeler, \&
  Vinko}]{Chatzopoulos2012}
Chatzopoulos, E., Wheeler, J.~C., \& Vinko, J. 2012, Astrophysical Journal,
  746, 121, \dodoi{10.1088/0004-637X/746/2/121}

\bibitem[{{Chevalier}(1982)}]{1982ApJ...258..790C}
{Chevalier}, R.~A. 1982, \apj, 258, 790, \dodoi{10.1086/160126}

\bibitem[{{Eddington}(1926)}]{1926ics..book.....E}
{Eddington}, A.~S. 1926, {The Internal Constitution of the Stars}

\bibitem[{{Kasen} \& {Bildsten}(2010)}]{2010ApJ...717..245K}
{Kasen}, D., \& {Bildsten}, L. 2010, \apj, 717, 245,
  \dodoi{10.1088/0004-637X/717/1/245}

\bibitem[{{Liu} {et~al.}(2018){Liu}, {Zhang}, {Wang}, \&
  {Dai}}]{2018ApJ...868L..24L}
{Liu}, L.-D., {Zhang}, B., {Wang}, L.-J., \& {Dai}, Z.-G. 2018, \apjl, 868,
  L24, \dodoi{10.3847/2041-8213/aaeff6}

\bibitem[{{Matzner} \& {McKee}(1999)}]{1999ApJ...510..379M}
{Matzner}, C.~D., \& {McKee}, C.~F. 1999, \apj, 510, 379,
  \dodoi{10.1086/306571}

\bibitem[{Moriya {et~al.}(2013)Moriya, Maeda, Taddia, Sollerman, Blinnikov, \&
  Sorokina}]{10.1093/mnras/stt1392}
Moriya, T.~J., Maeda, K., Taddia, F., {et~al.} 2013, Monthly Notices of the
  Royal Astronomical Society, 435, 1520, \dodoi{10.1093/mnras/stt1392}

\bibitem[{Nagy {et~al.}(2014{\natexlab{a}})Nagy, Ordasi, VinkÃ³, \&
  Wheeler}]{Nagy2014}
Nagy, A.~P., Ordasi, A., VinkÃ³, J., \& Wheeler, J.~C. 2014{\natexlab{a}},
  Astronomy and Astrophysics, 571, 77, \dodoi{10.1051/0004-6361/201424237}

\bibitem[{Nagy {et~al.}(2014{\natexlab{b}})Nagy, Ordasi, Vinkš®, \&
  Wheeler}]{Nagy_2014}
Nagy, A.~P., Ordasi, A., Vinkš®, J., \& Wheeler, J.~C. 2014{\natexlab{b}},
  Astronomy \& Astrophysics, 571, A77, \dodoi{10.1051/0004-6361/201424237}

\bibitem[{{Nagy} \& {Vink{\'o}}(2016)}]{2016A&A...589A..53N}
{Nagy}, A.~P., \& {Vink{\'o}}, J. 2016, \aap, 589, A53,
  \dodoi{10.1051/0004-6361/201527931}

\bibitem[{{Popov}(1993)}]{1993ApJ...414..712P}
{Popov}, D.~V. 1993, \apj, 414, 712, \dodoi{10.1086/173117}

\bibitem[{Wang {et~al.}(2016)Wang, Wang, Dai, Xu, Han, Wu, \& Wei}]{Wang_2016a}
Wang, L.-J., Wang, S.~Q., Dai, Z.~G., {et~al.} 2016, The Astrophysical Journal,
  821, 22, \dodoi{10.3847/0004-637x/821/1/22}

\bibitem[{{Zampieri} {et~al.}(2003){Zampieri}, {Pastorello}, {Turatto},
  {Cappellaro}, {Benetti}, {Altavilla}, {Mazzali}, \&
  {Hamuy}}]{2003MNRAS.338..711Z}
{Zampieri}, L., {Pastorello}, A., {Turatto}, M., {et~al.} 2003, \mnras, 338,
  711, \dodoi{10.1046/j.1365-8711.2003.06082.x}

\end{thebibliography}

\end{document}